# Comet and Meteorite Traditions of Aboriginal Australians

Duane W. Hamacher
Nura Gili Centre for Indigenous Programs, University of New South Wales,
Sydney, NSW, 2052, Australia
Email: d.hamacher@unsw.edu.au

Of the hundreds of distinct Aboriginal cultures of Australia, many have oral traditions rich in descriptions and explanations of comets, meteors, meteorites, airbursts, impact events, and impact craters. These views generally attribute these phenomena to spirits, death, and bad omens. There are also many traditions that describe the formation of meteorite craters as well as impact events that are not known to Western science.

**Comets**

Bright comets appear in the sky roughly once every five years. These celestial visitors were commonly seen as harbingers of death and disease by Aboriginal cultures of Australia. In an ordered and predictable cosmos, rare transient events were typically viewed negatively – a view shared by most cultures of the world (Hamacher & Norris, 2011). In some cases, the appearance of a comet would coincide with a battle, a disease outbreak, or a drought. The comet was then seen as the cause and attributed to the deeds of evil spirits. The Tanganekald people of South Australia (SA) believed comets were omens of sickness and death and were met with great fear. The Gunditjmara people of western Victoria (VIC) similarly believed the comet to be an omen that many people would die. In communities near Townsville, Queensland (QLD), comets represented the spirits of the dead returning home. The coincidence of a comet and a drought led the Euahlayi people of New South Wales (NSW) to believe comets were evil spirits that sucked the water from clouds, causing the drought. The tail of the comet was the large family that consumed the water from the river.

Other Aboriginal groups gave significance to the tail of a comet resembling smoke. Aboriginal communities of the Talbot district (VIC), Cape York (QLD), and Bentinck Island in the Gulf of Carpentaria (QLD) made this association – one shared with the Maori of Aotearoa (New Zealand). Groups in the Central Desert of the Northern Territory (NT) believed comets represented celestial spears thrown by spirits or ancestors in the sky, including the Pitjantjatjarra, Kaitish, Luritja, and Arrernte.

**Meteors**

Views of meteors across Aboriginal cultures are many and diverse, but most relate to death and evil spirits (Hamacher, 2011; Hamacher & Norris, 2010). A frequent interpretation is that meteors were the spirits of the recently deceased. In Wardaman traditions (near Katherine, NT), when a person dies their spirit ascends into the sky and passes through the star *Garrndarin* (Spica), becoming a star of its own. It looked after by the Rock-Cod star *Munin* (Arcturus) before falling back to Earth as a meteor. The star-spirit fell into a stream, where the Earthly Rock-Cod looked after it again. The spirit then pursued a potential mother and entered her to be reincarnated as a baby.

Across northern Australia, meteors are seen as the fiery eyes of celestial serpents (sometimes called the Rainbow Serpent). The Tiwi people of Bathurst and Melville islands north of Darwin, NT see meteors as the *Papinjuwari* – evil spirits with long claws that steal the hearts of babies. In the Yolngu traditions of Arnhem Land (NT), a similar spirit is called *Namorrorddo*, and in Lardil traditions of Mornington Island (QLD) it is called *Thuwathu*.

Meteor traditions also involved war. Aboriginal groups in northern QLD. followed the trails of meteors, believing the falling stars would lead them to enemy warriors. The Ngarigo people of New South Wales believed a meteor pointed to the direction of a group preparing for war. The Wathi-Wathi people of the Murray River region in Victoria perceived meteors as the path of a nulla-nulla (a short spear) in the sky. Similarly, Arrernte revenge rituals in the Central Desert (NT) involved throwing a small spear filled with evil magic in the direction of the intended target. If a meteor appeared, it signified the death of the person.

**Impact Events**

Oral traditions of Aboriginal groups across Australia describe fiery stars falling from the sky with a roar, striking the land and causing death and destruction. All of these accounts, except for Henbury (described below), do not coincide with impact sites known to Western science (Hamacher & Norris, 2009).

The Weilan people of north-central NSW tell of a large star that fell to Earth, lighting up all of the surrounding land. A similar story is recounted from the Aboriginal people of Wilcannia in north-western NSW who describe a large fiery star that rumbled and smoked as it fell from the sky, crashing into the ground in the Darling riverbed northeast of Wilcannia at a place called *Purli Ngaangkalitji*. The impact was followed by a deluge. It is not certain if these stories are related, but no known impact sites are located in New South Wales.

Yuin traditions from the Shoalhaven region near Nowra, NSW (south of Sydney), describe an impacting meteor shower and airburst. In the story, the sky heaved and many stars fell to the Earth, flashing in the sky. A large reddish mass burst in the air, causing a deafening roar and scattered debris across the region, leaving burnt holes in the ground. A Gurudara story from east of Darwin, NT tells how a bright star named *Nyimibili* fell from the sky onto the camp near the Wildman River, burning all of the grass and trees.

Some published research suggests that impact rates are higher than currently predicted by scientists, and that large impacts in the ocean have occurred in recent times (<1000 years), causing huge tsunamis to strike coastal Australia and New Zealand. These claims are circumstantial and have been refuted, but remain a controversial topic in modern studies of cultural astronomy and geomythology.

**Meteorite Craters**

Aboriginal people have oral Traditions that relate to the Gosses Bluff crater (NT), Liverpool crater (NT), Wolfe Creek crater (WA), and the Henbury crater field (NT) (Hamacher & Goldsmith, 2014). The Liverpool crater, which is 1.6 km wide and formed over 500 millions years ago, is called *Yingundji* in the Kunwinjku language of eastern Arnhem Land. It is believed to be the nest of a giant catfish, and rock art depicting catfish were found in caves along the crater wall.

The Wolf Creek crater, located in the eastern Kimberley region of Western Australia, was formed 300,0000 years ago and is 900 m wide. It is called *Kandimalal* in the local Jaru language, and there are several stories about how the crater formed (Goldsmith, 2000). These include a being that was digging for yams, a Rainbow Serpent emerging from the ground, and a star falling to the Earth. Some traditions seem to have incorporated the views from Western scientists working in the area, as attested by an Elder who claimed the star-story was "white man's story" (Reeves-Sanday, 2007).

Gosses Bluff crater, west of Alice Springs in central Australia (NT), is 22 km wide and heavily eroded, having formed about 142 million years ago. What remains is a ring shaped mountain range, 5 km in diameter and 150 m in height. The Western Arrernte people call it *Tnorala*. In Arrernte traditions, a group of eight women were going to take the form of stars and have a corroboree in the Milky Way. One of the women set her baby in a wooden basket called a turna. As she danced in the ceremony with the other women, the baby rolled off the Milky Way and fell to the Earth. The impact of the baby and the turna drove the rocks upward, creating the ring-shaped mountain range. The turna can be seen in the sky as the constellation Corona Australia (the Southern Crown).

One of the youngest impact sites in Australia is the Henbury crater field, south of Alice Springs. It formed when a meteoroid broke apart in the atmosphere and created 13 craters covering a square kilometer. The impact occurred < 4,700 years ago and was probably witnessed by Aboriginal people. Oral traditions collected in the 1930s describe the crater field as a place where a fire-devil came from the sun and ran down to the Earth, creating the craters. The spirit burned and ate the people for breaking sacred laws. Aboriginal people would not collect water in the craters for fear the fire-devil would fill them with iron.

**Meteorites**

Some Indigenous people used meteorites as axe heads, but were generally viewed with reverence as sacred objects. Aboriginal people were "deadly afraid" of meteorites from the Tenham meteorite fall of 1879 in western Qld. A large exposed meteorite outside of Melbourne was struck with the axes of some local Aboriginal people. There is evidence of Aboriginal people transporting meteorites long distances, but the reasons for this are not known (Bevan & Bindon, 1996). In the early 20[th] century, interest in meteorites by White Australians prompted Aboriginal people to collect meteorites and tektites for sale to interested parties. Tektites were commonly used as surgical tools and ritual implements (Baker, 1957; Edwards, 1966).

**References**

Baker, G. (1957). The Role of Australites in Aboriginal Customs. *Memoirs of the National Museum of Victoria*, 22(8), 1–26.

Bevan, A. W. R., & Bindon, P. (1996). Australian Aborigines and meteorites. *Records of the Western Australian Museum*, 18, 93-101.

Edwards, R. (1966). Australites Used for Aboriginal Implements in South Australia. *Records of the South Australian Museum*, 15(2), 243–251.

Goldsmith, J. (2000). Cosmic Impacts in the Kimberley. *Landscope Magazine*, 15(3), 28-34.

Hamacher, D. W. (2011). Meteoritics and cosmology among the Aboriginal cultures of Central Australia. *Journal of Cosmology*, 13, 3743-3753.

Hamacher, D. W. and Goldsmith, J. (2013). Aboriginal oral traditions of Australian impact craters. *Journal of Astronomical History and Heritage*, 16(3), 295-311.

Hamacher, D. W. and Norris, R. P. (2011). Comets in Australian Aboriginal astronomy. *Journal of Astronomical History and Heritage*, 14(1): 31-40, 2011.

Hamacher, D. W. and Norris, R. P. (2010). Meteors in Australian Aboriginal Dreamings. *WGN - Journal of the International Meteor Organization*, 38(3), 87-98.

Hamacher, D. W. and Norris, R. P. (2009). Australian Aboriginal geomythology: eyewitness accounts of cosmic impacts? *Archaeoastronomy*, 22, 60-93.

Reeves-Sanday, P. (2007). *Aboriginal Paintings of the Wolfe Creek Crater*. Philadelphia: University of Pennsylvania Museum of Archaeology and Anthropology.